\documentstyle[aps,twocolumn,prl,epsf]{revtex}

\newcommand{\beq}{\begin{eqnarray}}
\newcommand{\eeq}{\end{eqnarray}}

\begin{document} 
\twocolumn[\hsize\textwidth\columnwidth\hsize\csname
@twocolumnfalse\endcsname

\title{Evolutionary dynamics in the Bak-Sneppen model on small-world
       networks} 
\author{R. V. Kulkarni\cite{e-mail}, E. Almaas and D. Stroud}
\address{Department of Physics, The Ohio State University,
         Columbus, OH 43210} 

\date{\today}

\maketitle

\begin{abstract} 
We study the dynamics of the Bak-Sneppen model on small-world
networks. For each site in the network, we define a ``connectance,''
which measures the distance to all other sites. We find radically
different patterns of activity for different sites, depending on their
connectance and also on the topology of the network. For a given
network, the site with the minimal connectance shows long periods of
stasis interrupted by much smaller periods of activity. In contrast,
the activity pattern for the maximally connected site appears uniform
on the same time scale. We discuss the significance of these results
for speciation events.
\end{abstract}

\draft 
\vskip1.5pc]

\par 

The theory of punctuated equilibrium(PE) \cite{GE} states that (i)
most evolutionary change associated with life on Earth occurs during
speciation events, and (ii) the time scale for these speciation events
is very brief compared to the lifetimes of the individual
species. Thus the evolutionary history of most species is
characterized by long periods of stasis punctuated by relatively brief
intervals of rapid evolutionary activity. Mayr's theory of allopatric
speciation\cite{Mayr}, which forms the basis of PE, suggests that the
rapid evolutionary activity leading to speciation generally occurs in
small isolated populations, whereas more widespread populations
exhibit little evolutionary change over the same time scale.
Understanding why the rates of evolutionary change differ so
drastically in these two types of populations is one of the important
problems of evolutionary biology. A related problem is the study of
patterns of extinctions in biological history, which seem to show
scale-free behavior as discussed by Raup\cite{Raup}. Stanley
\cite{Stanley} has argued that these problems, and most large-scale
trends in evolutionary history, can be understood by using the species
as a fundamental unit of evolution.

\par 

Bak and Sneppen \cite{BS} have introduced a simple model aimed at
understanding these evolutionary patterns using the species as the
fundamental unit. The model is based on coevolution of species and
exhibits ``intermittent dynamics'' - that is, species undergo long
periods of little change, called stasis, which are punctuated by
sudden bursts of activity called avalanches. It provides a natural
explanation for the apparent scale-free behavior of the extinctions -
the system evolves into a self-organized critical (SOC) state with
avalanches (which are correlated with extinction events) occurring at
all scales. The same model can also be used to understand the
scale-free behavior of seemingly unrelated phenomena such as
earthquakes, as noted by Ito\cite{Ito}. Aside from its applications to
various problems, the Bak-Sneppen model is of intrinsic interest,
since it is one of the simplest models giving rise to SOC behavior.

\par

The Bak-Sneppen model, as elaborated below, has been extensively
studied for regular networks \cite{PMB}. However, as argued by Watts
and Strogatz\cite{WS}, most real-life networks are neither perfectly
ordered nor completely disordered but fall under the category of
``small-world'' networks which interpolate smoothly between the two
extremes. Such networks are characterized by a high degree of local
order, yet appear disordered on a large-scale because of the presence
of shortcuts in the networks. Because of their wide applicability,
there have recently been numerous papers characterizing the properties
of such networks\cite{cond-mat}. In the context of species interacting
in an ecosystem, examples of food webs indicate that the interactions
are better represented by small-world networks rather than by a simple
ordered topology.

\par

The aim of this Letter is to examine the dynamics of the Bak-Sneppen
model on small-world networks. In our analysis, we find it useful to
define a site-dependent property which we call ``connectance.'' We
find that the patterns of activity at each site are correlated with
the connectance. In particular, the minimally connected site in a
given network shows activity characteristic of intermittent dynamics,
whereas the maximally connected site shows uniform activity on the
same time scale. Furthermore, for the maximally connected site we see
a drastic reduction in the stasis times in going from the ordered
topology to the small-world networks. We discuss the implications of
these results for speciation events.

\par

We begin by recalling the definition of the Bak-Sneppen model. Each
species is represented by a site on a 1-$d$ lattice with periodic
boundary conditions. Each lattice site is connected to its $2k$
nearest neighbors by a bond (so that for $n$ interacting sites, or
species, we have $nk$ bonds). With each site we associate a random
number, called the barrier value, between 0 and 1; this number plays
the role of a barrier against evolutionary change for that
species. The dynamics of the Bak-Sneppen model is carried out by the
following rule. At each time step, we select the site with the {\em
minimal} barrier value (denoted the minimal site), and assign new
barrier values to this site and its $2k$ neighboring sites. We refer
to a reassignment of the barrier value at a site as activity at that
site. This set of rules, called extremal dynamics, leads the system
into a SOC state where the distribution of barrier values is uniform
above a critical barrier value $f_c$ \cite{PMB}.

\par

The small-world networks are generated using the procedure outlined by
Watts and Strogatz \cite{WS}. We start with the 1-$d$ lattice
described above and ``rewire'' each bond with a probability $p$ ($0 \le
p \le 1$). The rewiring consists of taking a given bond and moving it so
that, instead of connecting two neighbors, one end of the bond is
moved so as to connect with another site chosen at random, with the
constraint that double bonds are forbidden. The network so formed can
then be characterized by the functions $L(p)$, which is defined as the
number of bonds in the shortest path between two sites, averaged over
all pairs of sites, and $C(p)$ which is a measure of clustering in the
graph \cite{WS}. The network is said to fall in the small-world regime
if it satisfies the conditions $C(p)/C(0) \sim 1$ and $L(p)/L(0) \sim
0$. Table \ref{table1} shows the values of these two ratios for the
networks we studied, all of which were characterized by $n = 2000$ and
$k = 2$. A recent study on food webs \cite{Sugihara} suggests that our
choice of $k$ is often realized in practice. Finally, for each site,
we define the connectance $g_i$ by
\begin{eqnarray} 
	g_i &=& 1-\frac{D_i-\min _{\{j\}}D_j}{\max _{\{j\}}D_j-
	\min _{\{j\}}D_j} ~;~~p>0 \\ 
	D_i &=& \sum_{j=1}^{N} d(i,j)
\end{eqnarray}
where $d(i,j)$ is the minimal distance between sites $i$ and $j$, that
is, the minimal number of bonds which must be traversed in going from
$i$ to $j$.

\par

We carried out our simulations for $n=2000$ and for $p=0.0$, $0.01$,
$0.04$, and $1.0$. For each disordered network ($p \neq 0$) we studied
the dynamics for 10 different realizations. Remarkably, even though
the topologies for a given $p$ value were distinct, our calculated
quantities all collapsed onto the same set of curves. The results we
present were obtained by sampling for $5 \times 10^{9}$ time steps
after the stationary state characteristic of SOC had been reached. In
the following, we discuss results for $p=0.0$ and $p=0.01$ since these
capture the essential features in going from ordered to small-world
networks.

\par 

In the Bak-Sneppen model, an avalanche is defined as the sequence of
time steps for which the minimal site has

\begin{table}
\caption{The functions L(p)/L(0) and C(p)/C(0) as defined in the text,
calculated for a system with $n=2000$ sites and $k=2$. For each p, we
averaged over 25 different realizations.}

\begin{tabular}{l|l|l} 
p & L(p)/L(0) & C(p)/C(0)\\ 
\hline 
 0.01 & 0.14 & 0.97 \\ 
 0.04 & 0.06 & 0.89 \\ 
 1.00 & 0.02 & 0.04 \\ 
\end{tabular}
\label{table1}
\end{table}

\begin{figure}[tb]
\epsfysize=7cm
\centerline{\epsffile{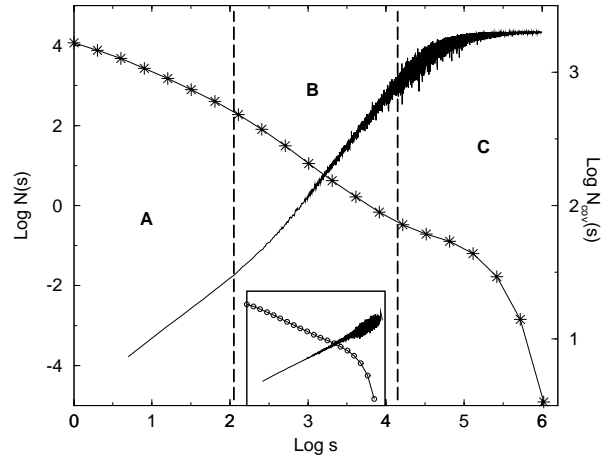}}
\caption{Number of avalanches ($\ast$) and number of sites covered
(full line without symbols; $\mbox{N}_{cov}$) by an avalanche
vs. duration $s$ of the avalanche, for $p=0.01$. The influence of
short-cuts, or small-world behavior, is reflected in the slope changes
between regions {\bf A} and {\bf B}. In region {\bf C}, finite-size
effects dominate. The inset shows the same plots for $p=0.0$.}
\label{Fig:aval}
\end{figure}

\noindent
a barrier value smaller than a threshold value $f_0$. The avalanche
thus ends when the minimal site has a barrier value greater than
$f_0$. For each $p$-value, we choose $f_0$ such that $\Delta
f=f_c-f_0=0.01$. Fig.\ \ref{Fig:aval} shows the distribution of
avalanche durations and the average number of sites covered
($\mbox{N}_{cov})$ by an avalanche of a given duration for $p = 0.0$
and $p=0.01$. For $p=0.0$ (inset) we see the expected power-law
behavior with a cut-off for large avalanche sizes. For $p=0.01$, on
the other hand, we see novel features arising from the small-world
properties of the network. In particular, we now see {\it two}
power-law regimes, denoted by regions {\bf A} and {\bf B}. As the
avalanche size grows, it increasingly ``senses'' the short-cuts in the
system, and becomes more delocalized; this delocalization produces an
increased slope in $\mbox{N}_{cov}$ in going from regime {\bf A} to
regime {\bf B}. A more detailed analysis of this behavior will be
presented elsewhere\cite{EKS}.

\par

\begin{figure}[tbp]
\epsfysize=7cm
\centerline{\epsffile{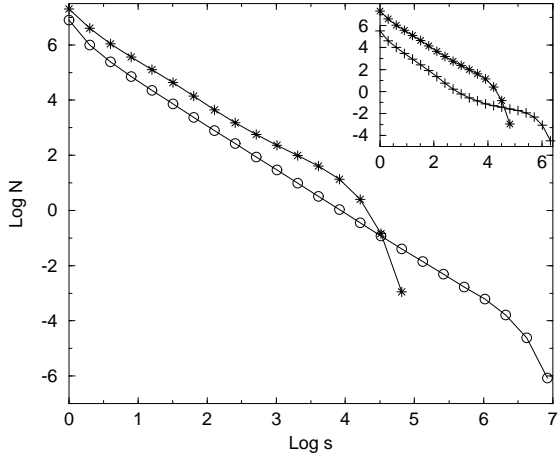}}
\caption{Distribution of first-return (stasis) times for the maximally
connected sites for $p=0.0$ ($\circ$) and $p=0.01$ ($\ast$). Note that
the cut-off, which corresponds to the maximal stasis time, is reduced
by 2 orders of magnitude in going from $p=0.0$ to $p=0.01$. The inset
shows the distribution of stasis times for the maximally ($\ast$) and
the minimally ($+$) connected site in a $p=0.01$ network.}
\label{Fig:first}
\end{figure}

In Fig. \ref{Fig:first}, we show the so-called first-return plots,
i.e., the distribution of waiting times between subsequent returns of
activity, at the maximally connected sites for $p=0.0$ and $0.01$.
All the plots show power-law behavior followed by a cut-off which
corresponds to the maximal stasis time. We note that the maximal
stasis time decreases by roughly two orders of magnitude in going from
$p=0.0$ to $p=0.01$. Thus the transition from ordered to small-world
networks is marked by a drastic reduction of the maximal stasis times.
The inset shows the first-return plots for the maximally and the
minimally connected sites in a $p=0.01$ network. From these plots we
see that the cut-off is correlated with the connectance: higher
connectance leads to a smaller cut-off time for a given network.

\par 

A dramatic consequence of the reduction in stasis times 

\begin{figure}[tbp]
\epsfysize=7cm
\centerline{\epsffile{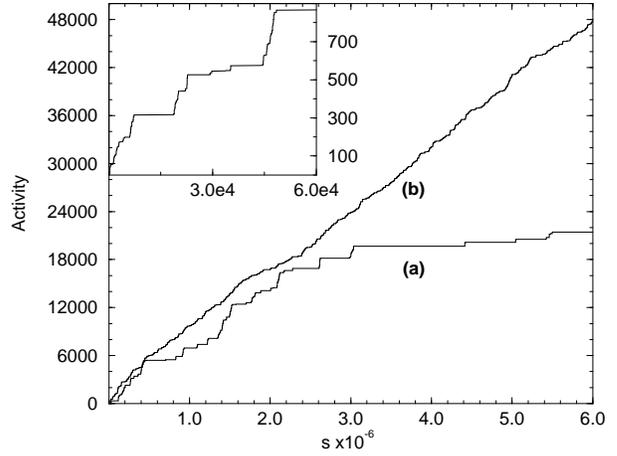}}
\caption{Calculated cumulative activity at the maximally connected
sites for (a) $p=0.0$ and (b) $p=0.01$. While (a) shows a ``punctuated
equilibrium'' pattern, (b) appears uniform on the same time scale. The
inset shows that actually (b) also exhibits a ``punctuated
equilibrium'' pattern, but on a much reduced time scale.}
\label{Fig:devil3}
\end{figure}

\begin{figure}[tbp]
\epsfysize=6.7cm
\centerline{\epsffile{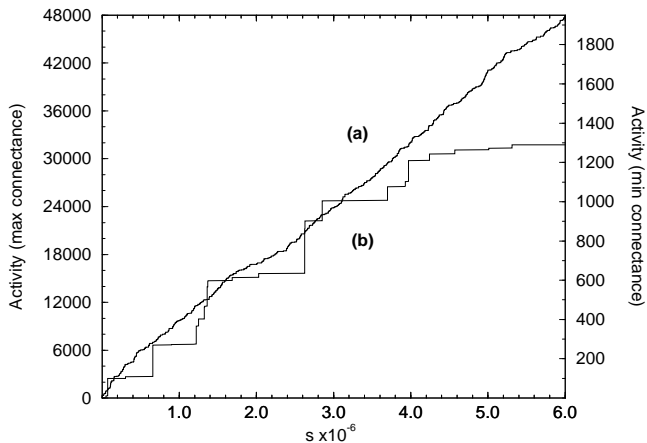}}
\caption{Cumulative activity for sites with (a) maximal and (b)
minimal connectance for $p=0.01$. The maximally connected site has a
cumulative activity which is $\sim 40$ times that of the minimally
connected site.}
\label{Fig:devil}
\end{figure}

\noindent
can be seen in Fig.\ \ref{Fig:devil3}, where we show the activity
plots at the maximally connected sites for $p=0.0$ and $p=0.01$. For
$p=0.0$, the cumulative activity at a particular site shows a pattern
of ``punctuated equilibrium:'' long periods of stasis interrupted by
much shorter periods of activity. By contrast, the maximally connected
site at $p >0$ shows a {\em uniform} pattern of activity on the same
time scale. This behavior results from the decrease in maximal stasis
time (cf.\ Fig. \ref{Fig:first}) with increasing $p$. This
relationship is clearly indicated in the inset, which shows that the
same pattern now exhibits ``punctuated equilibrium'' behavior on a
much smaller time scale. The correlation between maximal stasis time
and connectance for a given network can be seen in
Fig.\ \ref{Fig:devil}, where the activity is plotted for the sites with
maximal and minimal connectance at $p=0.01$. The minimally connected
site shows a pattern of ``punctuated equilibrium,'' whereas the site
with maximal connectance exhibits uniform behavior. In short, the
periods of stasis are significantly reduced when either $p$ or $g$ is
increased.

\par 

The uniform activity pattern seen for $p > 0$ may appear to be at
variance with the theory of punctuated equilibrium. But in fact, we
argue that this behavior actually supports that theory in the form
postulated by Gould and Eldredge \cite{GE}. We first clarify the
correspondence between our idealized model and an ecosystem. As
already noted, the sites correspond to species, while the network of
connections between them can be viewed as representing the food web of
the ecosystem. Activity at a site represents a significant change for
the corresponding species population in the ecosystem; an accumulation
of activity can lead to either speciation or extinction. Our

\noindent
model does not distinguish between the two, and we expect instances of
both speciation and extinction to be proportional to the amount of
activity. Finally we note that a bond between two species represents a
dependency link between them, which is typically, but not exclusively,
a predator-prey relationship.

\par

We now discuss the correspondence between the activity patterns
described above and the theory of punctuated equilibrium. We note that
the staircase pattern seen in Fig.\ \ref{Fig:devil3}(a) corresponds to
an evolutionary process called anagenesis, in which the entire
ancestral species evolves into a new species. But, as noted by Stanley
\cite{Stanley}, most speciation processes in nature are actually {\em
branching} processes (called cladogenesis), in which a new species is
created from a geographically isolated sub-population of the ancestral
species. During cladogenesis, the ancestral species shows little
change (stasis), whereas the isolated sub-population undergoes rapid
evolutionary change. This is the basic process involved in the theory
of punctuated equilibrium. Thus, what is required is that the periods
of stasis be greatly reduced for the isolated population, in
comparison to the ancestral species. This behavior is precisely what
is seen in Fig.\ \ref{Fig:devil3}: in these plots, increasing $p$
results in much shorter periods of stasis and hence, an activity plot
which has the appearance of uniformity in time.

\par

Thus, our scenario for the time sequence of allopatric speciation
events, as motivated by our simulations, is the following. (1) A
sub-population migrates to a peripherally isolated region, such as an
island. We expect the island ecology to have a higher value of $p$ for
the interaction network; hence, the periods of stasis for the isolated
sub-population should be much smaller than those of the

\noindent
ancestral population. (2) The migration event itself initiates a
period of activity on the island \cite{Brown}, while stasis continues
to prevail for the ancestral population. The combination of these two
effects gives rise to rapid evolutionary change leading to speciation
for the island population, and simultaneously stasis in the ancestral
population.

\par

Note that this model does not exclude other proposed mechanisms, such
as the ``founder effect'' proposed by Mayr\cite{Mayr}, but instead is
complementary to them. Our scenario focuses on the coevolutionary
activity in different ecosystems and thus on effects {\em external} to
the species populations, whereas the other mechanisms focus on the
internal characteristics of the species populations. For example, the
founder effect requires that the peripherally isolated population be
small, and therefore more amenable to rapid change. Speciation can,
however, occur rapidly over a large area containing many millions of
individuals. In particular, Williamson \cite{Williamson} has shown
that, in a group of African lakes in the Turkana basin, which shrank
and became separated from the parental water bodies, the species
underwent significant phylogenetic change over only $5 \times 10^{4}$
years, whereas the ancestral populations remained virtually unchanged
in their parental waters. In this case, the observed speciation events
cannot be explained by the founder effect, as argued by Williamson,
but they do seem consistent with our proposed scenario.

\par

Our calculations show that, for a given network, the patterns of
activity are strongly correlated with the connectance. How is this
observation reflected in studies of real ecosystems? One possibility
is the following: since the bonds in the network corresponds to
dependency links, the species with high connectance have a high degree
of dependency - i.\ e., they are specialists ({\it stenotypic}) -
whereas species with low connectance are generalists ({\it
eurotypic}). As already noted in Fig.\ \ref{Fig:devil}, the cumulative
activity is an order of magnitude larger for the maximally connected
site than for the one with minimal connectance. Based on our model, we
therefore expect {\it stenotypic} species to speciate or go extinct
more often than their {\it eurotypic} counterparts. This correlation
between ecological specialization and speciation and extinction rates
has been observed in numerous studies \cite{Eldredge}.

\par

In summary, we have extended the 1-$d$ Bak-Sneppen model to
small-world networks. This extension allows us to distinguish
different sites in the network, and thereby the species in an
ecosystem, based on a site-dependent property which we call
connectance. We find that the activity patterns are strongly
correlated with the topology of the network and with the
connectance. Finally, we discuss the possible implications of these
results for speciation events.

\par

We thank Jens O. Andersen for useful discussions. This work has been
supported by NASA through Grant NCC8-152 and through NSF through Grant
DMR97-31511. Computational support was provided by the Ohio
Supercomputer Center and by the San Diego Supercomputer Center.


\end{document}